\documentclass[a4paper]{article}

\usepackage{INTERSPEECH2018}

\title{Investigating accuracy of pitch-accent annotations in neural network-based speech synthesis and denoising effects }
\name{Hieu-Thi Luong$^1$, Xin Wang$^1$, Junichi Yamagishi$^1$, Nobuyuki Nishizawa$^2$}
\address{
  $^1$National Institute of Informatics, Tokyo, Japan 
  $^2$KDDI Research Inc., Saitama, Japan}
\email{\{luonghieuthi,wangxin,jyamagis\}@nii.ac.jp, no-nishizawa@kddi-research.jp}

\begin{document}

\maketitle
\begin{abstract}
We investigated the impact of noisy linguistic features on the performance of a Japanese speech synthesis system based on neural network that uses WaveNet vocoder. We compared an ideal system that uses manually corrected linguistic features including phoneme and prosodic information in training and test sets against a few other systems that use corrupted linguistic features. Both subjective and objective results demonstrate that corrupted linguistic features, especially those in the test set, affected the ideal system's performance significantly in a statistical sense due to a mismatched condition between the training and test sets. 
Interestingly, while an utterance-level Turing test showed that listeners had a difficult time differentiating synthetic speech from natural speech, it further indicated that adding noise to the linguistic features in the training set can partially reduce the effect of the mismatch, regularize the model, and help the system perform better when linguistic features of the test set  are noisy.
\end{abstract}

\section{Introduction}
Because of the rapid development of deep learning, more and more text-to-speech (TTS) synthesis systems adopt end-to-end approaches to some degree \cite{sotelo2017char2wav,wang2017tacotron,shen2017natural}. 
Although it has been reported that neutral-style synthetic speech from one system achieved a similar degree of quality and naturalness to natural recordings \cite{shen2017natural}, 
it is unknown how the end-to-end approach could perfectly avoid incorrect pronunciation \cite{shen2017natural} and make it possible to control prosody like the conventional structured architectures \cite{takashi2007style,watts2015sentence,luong2017adapting,henter2017principles}. 
More importantly, since most of the existing commercial TTS systems still adopt the pipeline structure which contains a front-end and a back-end, rapid shifting to a end-to-end architecture may be unable to answer how each part of the conventional structure contributes and limits the performance of existing TTS systems. Therefore, we believe that investigation on the pipeline of conventional TTS systems is still necessary and meaningful. In this work we adopted the conventional speech synthesis architecture which consists of three separate components: a linguistic analyzer, a neural network-based acoustic model \cite{zen2013statistical,ling2015deep} and a vocoder to synthesize waveforms from acoustic features.

As the initial step, our previous work \cite{wang2018comparison} has showed that the conventional TTS pipeline can be improved by replacing a deterministic vocoder \cite{kawahara2006straight,morise2016world} and RNN-based acoustic models \cite{fan2014tts,wang2017rnn} in the back-end with more advanced statistical models such as the WaveNet-based vocoder \cite{van2016WaveNet,ping2017deep}. 
However, our analysis revealed that the gap between synthetic speech and natural recordings still exists. One reason may be due to the fact that the statistical models in our previous work were trained by using linguistic feature automatically extracted from text. This motivated us to investigate the impact of the accuracy of the features on the back-end of the TTS system. There is a relevant investigation on the accuracy of phone sequences used for training of hidden Markov models \cite{7472660}. Our main focus in this paper is the accuracy of pitch accent information and a neural network. 

In this study, we first built an oracle system where manually corrected linguistic features were used for both model training and testing. Then, we compared the performance of the system with a few other systems that used corrupted linguistic features at training or/and testing stages. More particularly, we corrupted Japanese pitch accent types by adding discrete noise. From large-scale crowdsourcing listening tests, we found that in our neural network-based speech synthesis system, using corrupted linguistic features has a regularization effect (like a denosing auto-encoder) when linguistic features in the test set are noisy. We believe that this is a new finding in the speech synthesis field.

In section \ref{sec:systems} and \ref{sec:linguistic}, we describe the statistical models and linguistic features used in our TTS systems. In section \ref{sec:experiments}, we explain the methodology used to train and test our systems by using linguistic features with a varied amount of noise. In section \ref{sec:evaluations}, we list the results of both objective and subjective evaluation. Finally, in section \ref{sec:conclusions}, we discuss the findings and draw a conclusion.

\section{Speech synthesis back-end}
\label{sec:systems}
The back-end of the TTS system we investigated consists of two parts. The first part contains acoustic models that convert linguistic features into acoustic features such as the mel-generalized cepstral coefficients (MGC) and quantized fundamental frequency (F0). The second part is a WaveNet vocoder that generates speech waveform based on the basic of acoustic features.
All of the models adopt the configurations used in our previous work \cite{wang2018comparison}.

\subsection{Acoustic models}

\begin{figure}
\centering
  \includegraphics[width=0.95\columnwidth]{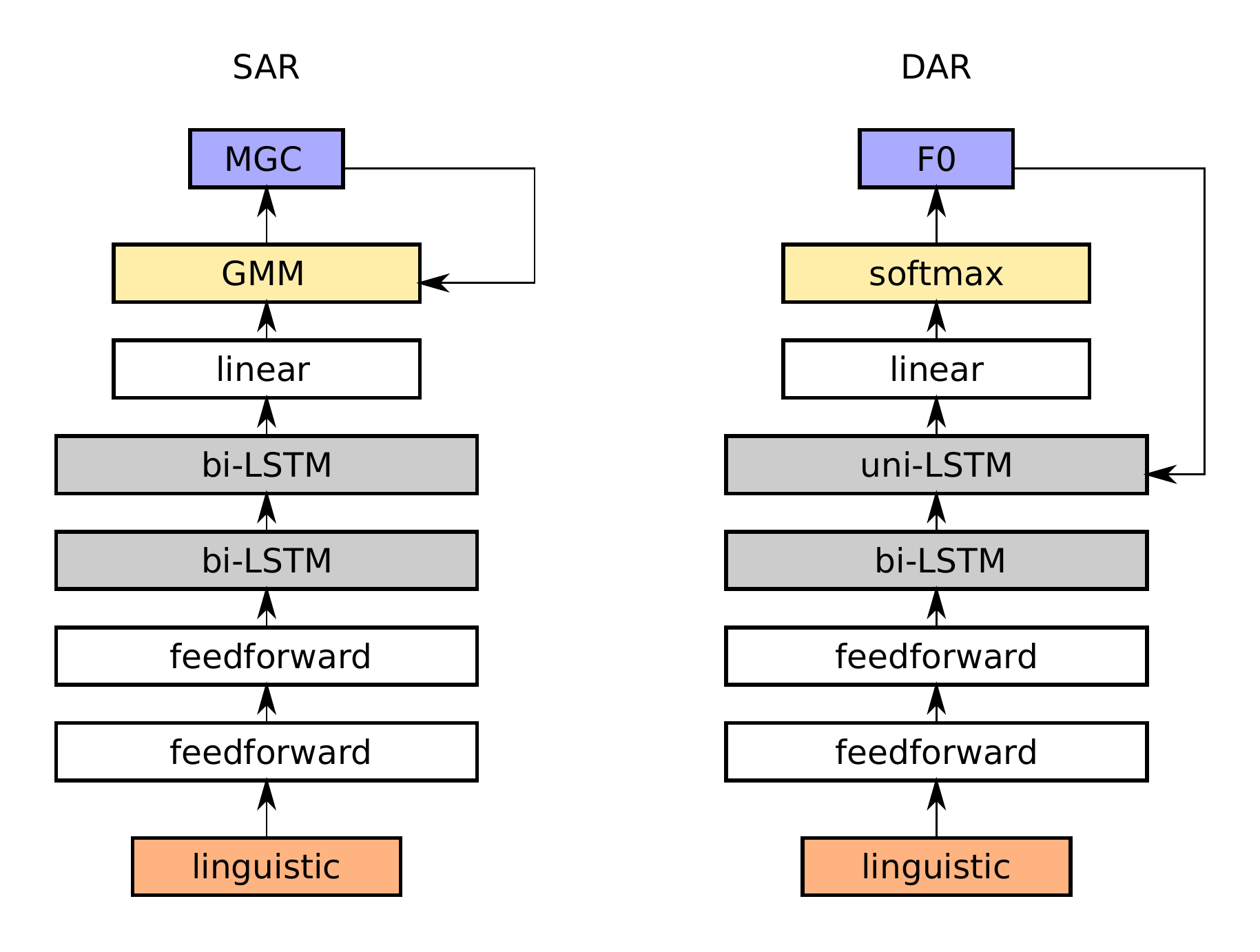}
  \vspace{-5mm}
  \caption{Structures of acoustic models used in our experiment. Feedforward layers use \textit{tanh} activation function while linear layers use linear activation function. GMM denotes Gaussian mixture models. Bi-LSTM and uni-LSTM denote bi-directional and uni-directional recurrent layers using long-short-term-memory (LSTM) units, respectively.}
  \label{fig:acoustic}
  \vspace{-5mm}
\end{figure}

The acoustic models are trained to learn the mapping from a sequence of linguistic features $\mathbf{l}_{1:N}=\{\mathbf{l}_1,\mathbf{l}_2,...,\mathbf{l}_N\}$ into a sequence of acoustic features $\mathbf{a}_{1:N}=\{\mathbf{a}_1,\mathbf{a}_2,...,\mathbf{a}_N\}$, where $N$ denotes the total number of frame. While a vanilla neural network can be used as the acoustic model, 
it assumes that $\{\mathbf{a}_1,\mathbf{a}_2,...,\mathbf{a}_N\}$ is a set of independent random variables given $\mathbf{l}_{1:N}$ even if convolution or recurrent layers are used. 
To overcome such weakness, we used autoregressive models, the basic idea of which is to feed the target data of the previous step as the input of the current step. 
On the basic of this idea, two separate autoregressive models plotted in Figure \ref{fig:acoustic} were trained to model MGC and quantized F0, respectively. 

The model for MGC was referred to as shallow autoregressive recurrent network (SAR). SAR maps a sequence of linguistics features to the value of a parameter set of which the distribution (in this case, a Gaussian distribution) of MGC of each frame can be specified. Different from a normal mixture density network \cite{bishop1994mixture}, SAR uses a linear function to summarize the acoustic features in previous frames and then changes the distribution of the current frame.
A similar network was used for quantized F0, which is referred to as deep autoregressive recurrent network (DAR). DAR was trained to map linguistic features to a quantized F0 representation rather than interpolated continuous-valued F0 data. Another distinct feature of DAR in comparison with SAR is that the output of the network is fed back to a recurrent layer that is closer to the input side.

The structure of the acoustic modelling networks are illustrated in Figure \ref{fig:acoustic}. Bi-directional and uni-directional long-short term memory (LSTM) layers were used after feedforward layers. Details on these models are given in our previous papers \cite{wang2017autoregressive,wang2017rnn}. 

\begin{table*}[t]
\caption{Definition and notations of TTS systems used in this experiment. Definition of linguistic feature sets is given in Section \ref{sec:data_feature}. \textbf{Oracle} + \textbf{Corrupted} means that 7999 randomly selected training utterances used corrupted labels, while rest used oracle linguistic features. Also shown are RMSE values and correlation of generated F0 and U/UV errors compared with natural speech. Mel-cepstral distortion (MCD) between generated and natural MGCs is also shown.}
\label{tab:objective}
\vspace{-2mm}
\centering
\scalebox{1.0}{
  \begin{tabular}{cccccccc}
  \hline
  \hline
    & \multicolumn{2}{c}{Linguistic features} & & \multicolumn{4}{c}{Objective measures} \\ 
    \cline{2-3}\cline{5-8}
   Notations & Train set & Test set &&  RMSE & CORR & V/U Error & MCD [dB] \\    \hline
   OJT & \textbf{OpenJTalk} & \textbf{OpenJTalk} & & 29.44 & 0.90 & 3.40\% & 4.72\\    
   MOO & \textbf{Oracle} & \textbf{Oracle} & & 23.31 & 0.94 & 3.25\% & 4.63 \\
   MOC & \textbf{Oracle} & \textbf{Corrupted} & & 31.09 & 0.89 & 3.29\% & 4.66\\
   MMC & \textbf{Oracle} +  \textbf{Corrupted} & \textbf{Corrupted} & & 28.15 & 0.91 & 3.20\% & 4.64\\
   MMO & \textbf{Oracle} +  \textbf{Corrupted} & \textbf{Oracle} & & 24.44 & 0.94 & 3.15\% & 4.63\\    \hline

  \end{tabular}
}
    \vspace{-5mm}

\end{table*}

\subsection{WaveNet vocoder}
To improve the quality of synthetic speech, we used a speaker dependent WaveNet vocoder. The WaveNet vocoder is a CNN-based autoregressive network that models a conditional distribution of a waveform sequence $\mathbf{o}_{1:T}$ over an auxiliary feature sequence $\mathbf{a}_{1:N}$ as
\begin{equation}
\centering
p(\mathbf{o}_{1:T}|\mathbf{a}_{1:N})=\prod\limits_{t=1}^{T}p(\mathbf{o}_{t}|\mathbf{o}_{<t},\mathbf{a}_{1:N}).
\end{equation}
For each sample $\mathbf{o}_t$ at time $t$, its value is conditioned on all of the previous observations $\mathbf{o}_{<t}$. In practice, the prediction of $\mathbf{o}_t$ was limited to a finite number of previous samples, which together were referred to as receptive field. By sequentially sampling the waveform per time step, the WaveNet vocoder can produce very high-quality synthetic speech in terms of naturalness, as reported in several papers \cite{van2016WaveNet,shen2017natural,ping2017deep}.

\section{Linguistic features used for our Japanese TTS system}
\label{sec:linguistic}

The linguistic features used for conventional Japanese TTS systems mainly include segmental and supra-segmental linguistic information. Despite the numerous differences in the two sets of linguistic features used in the experiments, i.e., OpenJTalk and oracle that will be introduced in Section 4.1, both sets contain quinphone contexts, word part-of-speech tags, pitch accent types of the accent phrases, interrogative phrase marks, and other structural information such as the position of the mora in a word, accent phrases, and utterances. These linguistic features will be used as the input of the acoustic model.

The two types of linguistic features that we were interested in for this investigation include the pitch accent type (Acc\_Type) and the interrogative phrase mark (Question\_Flag). The value of the pitch accent type is equal to the location of the accented mora in a Japanese accent phrase. It can also be a special number such as 0, which indicates a no-accent phrase. The interrogative phrase mark is binary and indicates whether a phrase is interrogative or not. These two types of features are essential to the prosody of Japanese utterances yet difficult to accurately obtain by using automatic prosodic annotation or text-analysis.

\section{Experiments}
\label{sec:experiments}
\subsection{Data and features}
\label{sec:data_feature}
This study used the same speech corpus as our previous work \cite{wang2018comparison}. This corpus has high-quality speech recordings of a female voice talent and was released as part of the Ximera datasets \cite{kawai2004ximera}.
Compared with our previous work, we excluded hundreds of utterances in which the manually annotated labels were unusable due to
imperfect pronunciation. This new training set contained 27,999 utterances while both the validation and test set contained 480 utterances. 
The duration of the training set was about 46.914 hours, among which the total silence at the two ends of the utterances was around 13.393 hours in total. 
The duration of the validation and test sets was 0.815 and 0.824 hours.

Acoustic features were extracted by using WORLD \cite{morise2016world} spectral analysis modules and SPTK. We used speech waveforms at a sampling frequency of 48 kHz to obtain these features with a window length of 25 ms and frame shift of 5 ms. 60-dimensional Mel-generalized cepstral coefficients (MGCs) and 25-dimensional band-limited aperiodicity values (BAP) were extracted. F0s were quantized into 255 levels as described in \cite{wang2017rnn}.
To investigate the impact of the accuracy of linguistic features, we prepared three sets of linguistic features:

\textbf{OpenJTalk:} the first set of linguistic features was extracted automatically from text by using OpenJTalk \cite{hts2015openjtalk}. These features were converted into 389 dimensional vector. This set is included as a reference because it was used in our previous work.

\textbf{Oracle:} the second set of linguistic features is based on in-house annotations provided by KDDI Research, Inc. The definition of the linguistic features is very similar to that used in the above first set, but it contains more precise phone definitions. Part-of-speech tagging is not included in the annotations. The dimension of the linguistic feature vector was 265. All features were manually verified.

\textbf{Corrupted:} the third set is based on the second set. However, we randomly changed the values of certain linguistic features. More specifically, we randomly added discrete noise ranging between -2 and +2 to the original value of [Acc\_Type] for each accent phrase with a 50\% probability. The value of the binary feature [Question\_Flag] for each accent phase was also randomly converted to the opposite value with a 30\% probability. We expected that these two types of processing would reproduce the annotation errors of Japanese accent types and question types.   

\subsection{Model configurations}
The structure of the acoustic models is plotted in Figure \ref{fig:acoustic}. The configuration of the layer size was 512 for feedforward, 256 for bi-directional LSTM-RNN, and 128 for uni-directional LSTM-RNN. The size of a linear layer depends on the size of the output. For SAR network, the output is a parameter set of Gaussian distributions for MGC, BAP, and voiced/unvoiced (V/UV) flags. BAP and V/UV were also included in the output even though they are not used to generate speech waveforms with the WaveNet vocoder. DAR used a similar configuration of layer size as SAR, but the output layer was a hierarchical softmax layer. 

Although the acoustic features were extracted from speech waveforms at a sampling frequency of 48 kHz, the WaveNet vocoder was trained by using speech waveforms at a sampling frequency of 16 kHz. PCM waveform samples were quantized into 10 bits after they were compressed by $\mu$-law coding \cite{modulation1988voice}. The network contained 40 causal dilated convolution layers similar to \cite{tamamori2017speaker}. WaveNet blocks were conditioned on MGC and quantized F0 parameters locally. The WaveNet vocoder was trained on acoustic features extracted from natural speech, while, in the generation stage, MGC and quantized F0 features predicted from DAR and SAR models were used. 

\subsection{Experimental conditions}
To investigate the impact of the noise in linguistic features, we trained a few systems by using different sets of linguistic features in the training and test stages as we described earlier. The definition and notations of each system can be found on the left part of Table \ref{tab:objective}. Note that the linguistic features for the validation set were not corrupted.
Also note that, instead of using the results of our previous study, we retrained the OpenJTalk-based model (OJT) by using the same data set configuration described in Section 4.1. Thus, the results of OJT can be compared with those of other experimental models.

\section{Evaluations}
\label{sec:evaluations}
\subsection{Objective evaluation}

Table \ref{tab:objective} shows the performance of each system in terms of RMSE, correlation, and V/UV errors of F0 trajectories converted from predicted F0 classes including an unvoiced class. As expected, the model trained and tested by using manually annotated labels, i.e., MOO, achieved the best results among the systems for all of the measurements. We can also see that when testing on the corrupted linguistic labels, the performance of MOC drastically dropped. 

Interestingly, MMC and MMO, which were trained by using partially corrupted linguistic features, performed better than MOC. For MMO, the objective results are comparable to those of MOO, which suggests that 7999 (around 28.57\% of all training data) corrupted labels did not affect the overall quality significantly. Meanwhile, MMC performed better than MOC even though MMC used corrupted linguistic features for training. Our hypothesis is that mixing corrupted labels with the training data is similar to a regularization method, such as the denoising auto-encoder and doing so helps a model generalize better and eases the negative impacts of the wrong information provided from incorrect linguistic features in the testing stage.

\subsection{Subjective evaluation}
The objective evaluation hinted at the performance of the acoustic models. However, because we used WaveNet vocoder and its sampling rather than a traditional deterministic vocoder to synthesize speech waveforms, it was necessary to test the overall quality of the synthetic speech samples subjectively. 

Therefore, we also conducted a large-scale subjective test.
Synthetic speech samples were generated by using the WaveNet vocoder. The natural speech was downsampled to 16 kHz and further converted to 8-bit $\mu$-law. Synthetic speech was converted to 8-bit and was normalized to have a similar volume to natural speech using the sv56 program \cite{p56}.

With the above five systems and the natural speech (NAT), each of which contained 480 utterances from the test set, we conducted two subjective tests. The first test was done to evaluate the mean-opinion-score (MOS) on a five point scale. The second was similar to a Turing test, where participants were asked to identify which of two samples presented is synthetic. During this test, an anchor question was included, where we presented the same natural speech twice. This question was expected to provide some insight into the nature of the testing environment. No default answers were given in any of these tests to make sure participants would have to make their own choices.

This large-scale listening test was conducted online through crowd-sourcing. Each participant was asked to navigate twelve pages for each set. Each page contained two questions, one for the MOS and another for the Turing test. The audio sample for the quality question contained a different sentence from that for the Turing question on the same page. 
One hundred subjects participated. They were allowed to repeat the test up to ten times. We collected a total of 720 sets, which led to 3 data points per unique audio sample for all of the systems.

\begin{figure}
\centering
  \includegraphics[width=0.92\columnwidth]{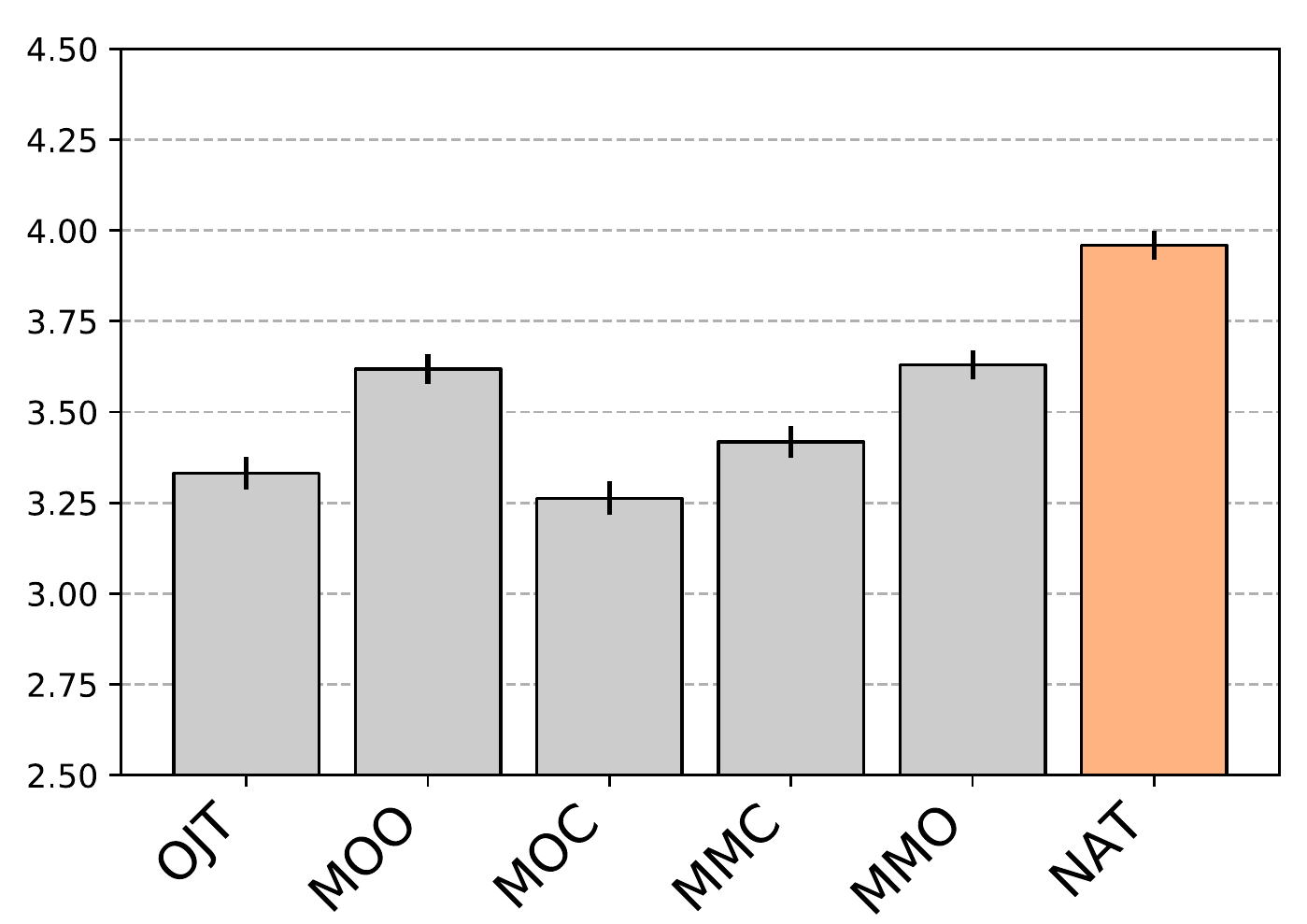}
  \vspace{-3mm}
  \caption{Subjective evaluation for quality of speech using MOS test. Bars indicate 95\% confidence intervals.}
  \label{fig:subjective}
    \vspace{-5mm}
\end{figure}

\textbf{Quality test:} Figure \ref{fig:subjective} shows subjective results for the quality test with a 95\% confidence interval with a student's t-distribution. Unsurprisingly natural speech still achieved the highest and most statistically significant score at 3.96 even when converted to the $\mu$-law encoding format. Audio samples generated by using manually annotated labels at the generation stage (MOO and MMO) achieved the second highest score, and the difference between MOO and MMO was not statistically significant (3.62 versus 3.63, p-value=0.720). We can also see that OJT and MOC performed the worst, and the difference between them was not statistically significant  (3.33 versus 3.26, p-value=0.05). Note that the p-value was calculated with Holm-Bonferroni correction.

What's interesting is that MMC, which used the corrupted linguistic features at both the training and testing stages, was better than OJT and MOC. These subjective results were consistent with the results of the objective evaluation on F0.

These results indicate a correlation between the accuracy of the linguistic features and the quality of the synthetic speech. A greater impact could be seen if the accuracy of annotated labels is high at the testing stage instead of the training stage.  Another finding is that, when linguistic features used in the test set contained noises, training the neural network models with a small amount of corrupted linguistic features seemed to improve the quality of the synthetic speech. We can also see that adding a small amount of corrupted linguistic features in the training set did not degrade the quality of the synthetic speech even if the test set did not contain any noise.

\begin{figure}
\centering
  \includegraphics[width=1.0\columnwidth]{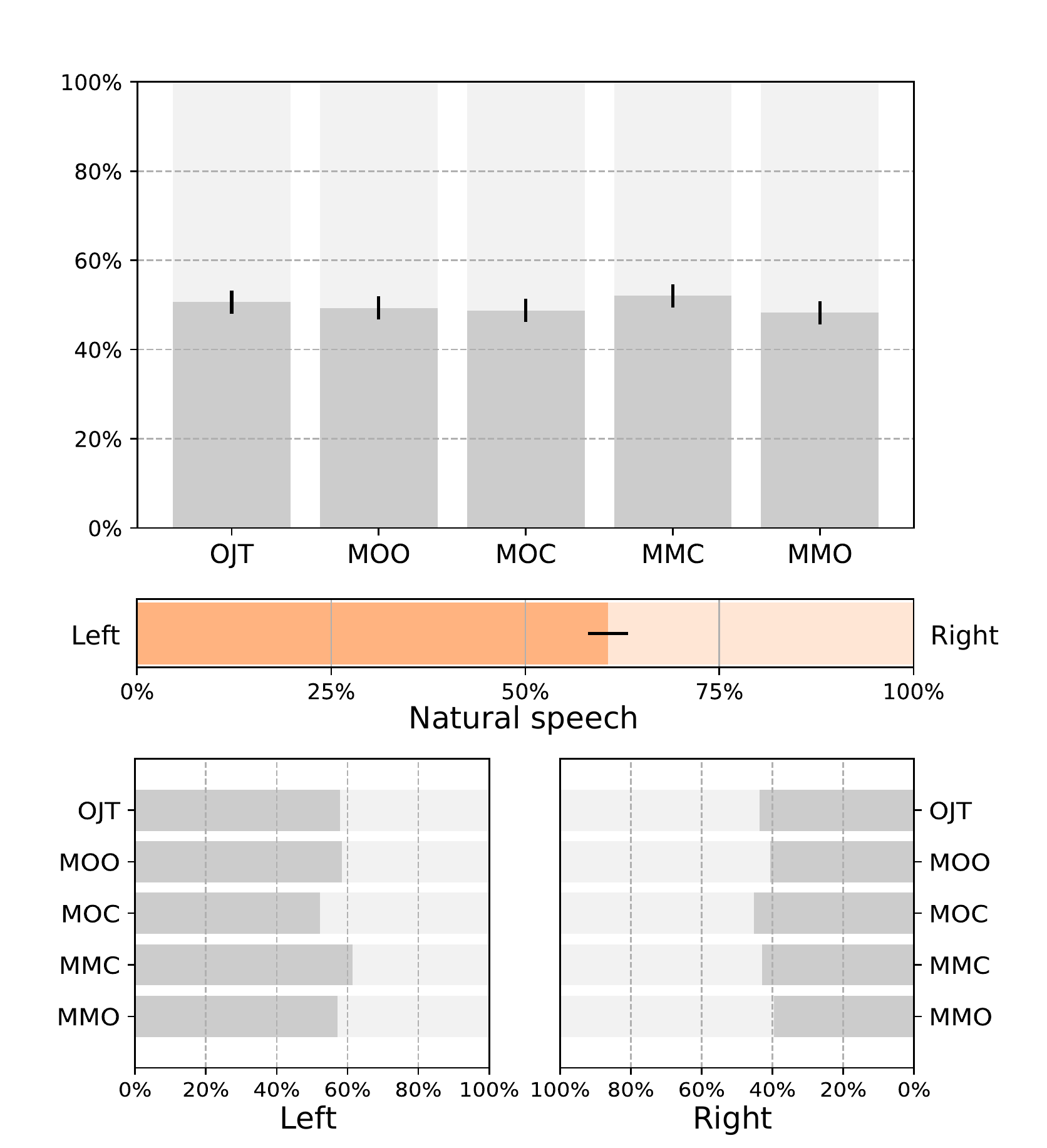}
  \vspace{-5mm}
  \caption{Results of utterance-level Turing (identification) test. Bars indicate 95\% confidence intervals. }
  \label{fig:turing}
    \vspace{-5mm}
\end{figure}

\textbf{Turing (Identification) test:} for the Turing test, participants were asked to identify which of two audio samples presented on left or right side of the web page was synthetic. The audio samples from one of the TTS systems and from natural speech were randomly switched between left and right to discourage subjects from developing any bias patterns. Figure \ref{fig:turing} shows the result. Surprisingly, for all comparisons between the synthetic and natural speech utterances, the correct-identification ratio was around 50\%, which suggested that our participants could not decide with certainty which of the two samples presented was synthetic. There was no significant difference between the five pairs of generated and (slightly degraded) natural speech. We think that this may not be surprising because the correlation of our F0 prediction model was as high as 0.9, we used a very large speaker-dependent corpus, that was larger than in a recent paper on Google's Tacotron 2 \cite{shen2017natural}, and natural speech was also slightly degraded by the $\mu$-law coding.

As we included an anchor test in our evaluation in which participants were asked to judge the differences between the same natural speech, it may be helpful to look into the result to gain some insight into our testing environment. 
The results we got for this anchor test showed that the left options were favored 60\% of the time, which suggested that participants had a slight bias for left option when it was difficult to choose the correct one. 
%
%
We can also analyze whether participants had a slight bias for left options for comparisons between the synthetic and natural speech utterances. Although the two options were randomly switched, from two sub figures at the bottom, we can see that the same tendency exists regardless of system types used. 
This basically gives some insight into developing a more sophisticated Turing test in the future.

With the outcomes for the Turing test, we can conclude that, while the synthetic speech did not achieve the same quality as natural speech, it was difficult for a normal human being to correctly determine the synthetic speech with our current state-of-the-art setups, at least when a reference natural-speech utterance was not offered.

\vspace{-3mm}
\section{Conclusions}
\label{sec:conclusions}
In this paper, we investigated the impact of noisy linguistic features on the performance of a Japanese speech synthesis system based on neural network that uses WaveNet vocoder. In this investigation, an ideal system that used manually corrected linguistic features in the training and test sets was compared against a few other systems that used corrupted linguistic features. The corrupted linguistic features, which were created by adding noises artificially to the correct pitch accent information. 

Both subjective and objective results demonstrate that corrupted linguistic features, especially those in the test set, affected our TTS system's performance significantly in a statistical sense due to mismatched conditions between the training and test sets. It was further indicated that adding noise to the linguistic features in the training set can partially reduce the effect of the mismatch, regularize the model, and help the system perform better when the linguistic features of the test set are noisy. As far as we know, this is a new finding in the speech synthesis field. Interestingly the utterance-level Turing test showed that our listeners had a difficult time differentiating synthetic speech from slightly degraded natural speech. 

Our future work includes comparing of our TTS system using manually corrected labels with recent end-to-end TTS systems and evaluating without using $\mu$-law coding.

\bibliographystyle{IEEEtran}
\bibliography{main}

\end{document}